\documentclass[times,final]{article}
\usepackage[left=2cm,right=2cm]{geometry}
\usepackage{color}
\usepackage{amssymb}
\usepackage{bm}
\usepackage{subcaption}
\usepackage{url}
\usepackage{graphicx}
\usepackage{amsmath} 
\usepackage{xcolor}

\begin{document}
	
	\begin{flushleft}
		{\Large\textbf\newline{Modeling straight and circle swimmers via immersed boundary methods: from single swimmer to collective motion.}
		}
		\newline
		\\
		Francesco Michele Ventrella\textsuperscript{1*},
		Guido Boffetta\textsuperscript{2},
		Massimo Cencini\textsuperscript{1},
		Filippo De Lillo\textsuperscript{1}.
		\\
		\bigskip
		{1} Dipartimento di Fisica and INFN, Universit\`a degli Studi di Torino, via P. Giuria 1, 10125 Torino.
		\\
		{2} Istituto dei Sistemi Complessi, CNR, via dei Taurini 19,00185 Rome, Italy and INFN, sez. Roma2 “Tor Vergata”.
		\\
		\bigskip
		* francescomichele.ventrella@unito.it
		
	\end{flushleft}

\begin{abstract}
We propose a minimal model of microswimmer based on immersed boundary methods.
We describe a swimmer (either pusher or puller) as
a distribution of point forces, representing the swimmer's flagellum and body, with only the latter subjected to no-slip boundary conditions with
respect to the surrounding fluid. In particular, our model swimmer consists of
only three beads (two for the body and one for the flagellum) connected by
inextensible and rigid links. When the beads are collinear, standard straight
swimming is realized and, in the absence of propulsion, we demonstrate that the swimmer's body
behaves as an infinitely thin rod.  Conversely, by imposing an angle between
body and flagellum the swimmer moves on circular orbits. We then discuss how
two swimmers, in collinear or non-collinear geometry, scatter upon encounter.
Finally, we explore the dynamics of a large number of swimmers
reacting to one another only via hydrodynamic interactions, and exemplify their
complex collective dynamics in both straight and circular swimmers.
\end{abstract}

\section{Introduction}
The study of motility in swimming animals and
microorganisms is a captivating subject in the biological realm,
encompassing various aspects such as feeding, reproduction and
prey-predator interactions
\cite{kiorboe2009mechanistic,elgeti2015physics} with potential
applications to biomedicine \cite{wang2021trends}.  Additionally, it
extends to the field of biological-inspired intelligent navigation
\cite{bonnard2018sub,colabrese2017flow}.  Moreover, in recent years, a
growing amount of research has focused on wet active
matter~\cite{marchetti2013hydrodynamics}, i.e. dense
suspensions of swimmers moving in a viscous fluid where the
hydrodynamic disturbances are a key mode of interaction.
Consequently, the dynamics of a single swimmer
becomes the focal point of numerous experimental
\cite{drescher2010direct,drescher2011fluid,di2010bacterial,carlson2020swimming},
theoretical \cite{lauga2009hydrodynamics}, and numerical
investigations
\cite{gustavsson2017finding,lauga2016bacterial,seyrich2018statistical,andersen2015quiet,chibbaro2021irreversibility}.
The overarching goal is to model the dynamics of a single swimmer in
its environment and understand how the interaction of these
organisms influences global behavior and the background flow field,
leading to collective organized motion
\cite{datt2019active,hoell2018particle,chibbaro2021irreversibility}.

Modeling self-propelled bodies can be broadly categorized based on
the streamlines they produce around them as ``pushers'' or ``pullers''
\cite{lauga2009hydrodynamics}.  Spermatozoa and some
  bacteria like {\em E. coli}, which propel via (single or bundled)
  flagella pushing the fluid away along the propulsion axis and drawing it
  in from the sides, are typical examples of pushers. Many
  biflagellates, such as microalgae like {\em Chlamydomonas}, which
  draw the fluid inwards along the propulsion axis and ejected
  it to the sides, are pullers.  Direct numerical
simulations are crucial to understand
  swimmer-swimmer and swimmer-fluid
interactions. Several models, with different degrees
  of complexity, have been developed, including the boundary integral
method for ellipsoidal swimmers
\cite{pozrikidis1992boundary,kanevsky2010modeling}, simple dumbbell
models
\cite{hernandez2005transport,hernandez2009dynamics,gyrya2010model,furukawa2014activity,cavaiola2021self},
Stokesian dynamics of 'squirmers' propelled by a surface slip velocity
\cite{ishikawa2007diffusion,ishikawa2008development}, immersed
boundary (IB) method
\cite{roma1999adaptive,peskin1995general,mittal2005immersed,lushi2013modeling}, penalty methods \cite{chibbaro2021irreversibility} and the
method of regularized Stokeslets for non-interacting swimmers
\cite{cortez2001method,cortez2005method,zhao2019method}.

This paper aims at proposing a swimmer model based on
immersed boundary methods. The IB method
\cite{peskin2002immersed}, initially developed to simulate blood flows
into the heart, has found applications in various biological fluid
dynamics problems
\cite{peskin1972flow,peskin1977numerical,peskin2020cardiac,peskin1980modeling,peskin1989three},
including animal locomotion
\cite{fauci1988computational,fauci1990interaction}.
In essence, the method treats the elastic material as part of the
fluid: body motion is obtained by interpolating the
forces due to fluid stress onto a set of points representing the
surface of the immersed body, and the body feedback
  on the fluid is applied by using the same interpolation
method. This allows the straightforward application of Navier-Stokes (NS)
solvers to complex flow geometries without the constraint of a
boundary-conforming grid, which is valuable especially in the case of
biological problems, where non-static walls or bodies are the norm.
In our case, the NS equations are solved using a
standard pseudo-spectral solver
\cite{fornberg1994review,gottlieb1977numerical,canutospectral,boyd2001chebyshev}
on a regular, triple-periodic grid, while each swimmer is represented
by as few as three Lagrangian points whose geometry is
prescribed by the internal forces.  Two kinds of swimmer
will be considered: a straight-swimming model in which the beads are
in a line and, in a still fluid, move in a straight line with a
stationary velocity proportional to a fixed propulsion force, also
parallel to the swimmer itself; a model in which the flagellum and the
body are at a constant angle, which at stationarity swims in a closed,
circular trajectory. We will call the second model a {\em circle
  swimmer}, based on previous literature
\cite{lowen2016chirality,ledesma2012circle,yang2014self,kummel2013circular,kaiser2013vortex}.

The paper is structured as follows. In Section II, we present the model and its numerical implementation, including a series of preliminary studies needed to set parameters and prove the robustness of the method. Sections III and IV presents the numerical results for single swimmers and a pair of swimmers in the straight and circular swimming mode, respectively. In Section V we present a preliminary exploration of the dynamics of a large number of swimmers and of how their collective organization changes when passing from straight to circular swimmers. The appendices present some more technical material: App.~A describes the Stokeslets solution used to fit the beads radius from the numerical simulations, App. B details the way inextensibility and rigidity are ensured in the model, while App. C presents an analytical derivation of the dynamics of the swimmer in the absence of propulsion, demonstrating that it behaves as an infinitely elongated rod.

\section{The immersed boundary method for a microswimmer}

\begin{figure}[ht!]
	\centering
\includegraphics[width=\textwidth]{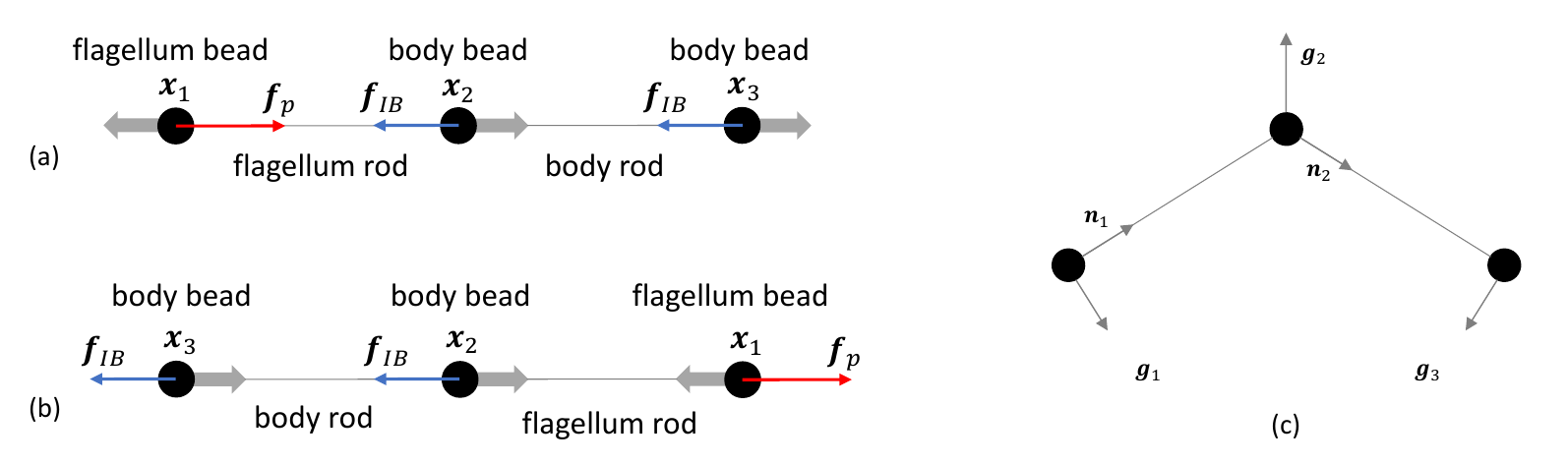}
\caption{Schematic
  view of the model swimmer. (a)
  Three-sphere model for a pusher. The red vector
  $\mathbf{f}_p$ is the propulsion force per unit of mass
  (i.e. acceleration) that allows the motion. The blue vector is the
  acceleration due to the no-slip condition. The gray arrows are the forces exerted by the pusher on the fluid.  (b)
  Three-sphere model for a puller, obtained from the
  pusher (a) by inverting the propulsion force. In both cases the
  swimming direction is from left to right. Notice
    that the position of beads $1,3$ is exchanged with respect to
    panel (a). (c) A generic configuration with an angle $\phi$
  between the body and the flagellum. The internal forces
  $\mathbf{g}_1,\mathbf{g}_2$ and $\mathbf{g}_3$ allow to control the angle or
  (like in this paper) keep it fixed. When $\phi\neq 0$ the swimmers
  perform a curvilinear motion.  }
\label{fig1}

\end{figure}

We consider a simple swimmer model consisting in a body and a
flagellum.  Following
\cite{cavaiola2021self,cavaiola2022swarm} both the
body and the flagellum are represented via a linear distribution of
spherical beads connected by inextensible
rods. Flagellum dynamics is not directly modeled and
the effects of propulsion are taken into account via localized forces
applied to the fluid.  The swimmer's body and flagellum are connected
by inextensible rods whose configuration is held constant by internal
forces, making the swimmer's shape rigid and inextensible. We will
consider both the case of straight swimmers, in which the flagellum is
parallel to the body, and that of circle swimmers, in
which flagellum and body are at a fixed angle, causing the swimmer
to move on a curved trajectory.  At each bead, a point force acts on
the fluid. The nature of the forces acting along the body differs
from those along the flagellum \cite{lushi2013modeling}.  The body is considered as a rigid
structure immersed in the fluid, along which no-slip conditions are
assumed for the fluid velocity. The no-slip conditions are
numerically enforced with a strategy derived from the IB methods,
which will be described shortly.  As a consequence, the body exchanges
momentum with the fluid through viscous interaction, with no further
modeling needed.  On the contrary, the flagellum beads are not
subject to no-slip conditions: they are instead used to apply the
propulsive force onto the fluid, while momentum conservation is
guaranteed by applying an opposite force on the beads themselves
and, thanks to rigidity of the rods, to the whole swimmer.

The simplest, bead-based swimmers proposed are made of two beads,
i.e. one for the body and one for the flagellum
\cite{cavaiola2021self}.  As discussed above, the flagellum bead is
not directly influenced by fluid velocity.  It follows that a
two-beads swimmer, with only one affected by the flow, is unaffected
by velocity gradients along its body and,
  consequently, cannot behave as a passive rod in limit of vanishing
propulsion.  The minimal swimmer must therefore have
at least two beads with no-slip boundary condition to
  describe the body.  In principle, one bead is sufficient to
describe the flagellum.  It was shown that, if the same number of
beads are used for the body and the flagellum, the velocity field
surrounding the swimmer in steady motion is qualitatively similar to
that produced by a force doublet \cite{cavaiola2021self}.  In the
following we will consider the simpler three-bead swimmer model, in
analogy to previous theoretical \cite{gyrya2010model} and numerical
works \cite{hernandez2005transport,furukawa2014activity}, which
studied similar models with slightly different approaches. One of
the novelties of the present paper is the possibility to
have curved trajectories, when the beads are not collinear. In
perspective, one can dynamically change the body-flagellum geometry
allowing for controlling the swimming direction. The latter
property can be exploited to model the dynamics of microrobots to be
employed, eg., in biomedical applications \cite{wang2021trends,
  li2017micro}
Figure~\ref{fig1}(a) represents a sketch of the three-sphere model for
a pusher. The flagellum bead is labeled as bead 1. The force acting
on it is the propulsion force per unit mass $\mathbf{f}_p$.  This force is
considered as fixed in modulus and parallel to the flagellum rod
connecting beads 1 and 2.  The inextensibility of the connecting rods
implies that a similar force is applied to the body beads so that, in
an otherwise still fluid, the resulting movement relative to the fluid
produces on the body beads two drag forces, denoted as $\mathbf{f}_{IB}$
in the figure, in the opposite direction.  Equal and opposite forces
(indicated in gray) are applied to the fluid in the corresponding
positions and guarantee momentum conservation.
The model for pullers (Fig.~\ref{fig1}(b)) is obtained by reversing  
$\mathbf{f}_p$ relative to the body. A generic, non-collinear configuration is
shown in Fig.~\ref{fig1}(c). We will show in Section IV that when the flagellum is at a fixed, non-zero angle with the body, the swimmer moves on a circular trajectory. For this reason we will refer to this case as a circle swimmer \cite{lowen2016chirality,ledesma2012circle,yang2014self,kummel2013circular,kaiser2013vortex}.

\subsection{The numerical implementation} 
\label{numericalmethod}
As outlined above, the swimmer is
described in terms of $N$ spheres with centers at the 
points $\mathbf{x}_i$, with $i=1,..,N$.
In what follows we will consider the cases $N=2$ and $N=3$ and 
assume that the 3D Eulerian problem of the evolution of
the velocity field $\mathbf{u}(\mathbf{x},t)$ is discretized in space on a uniform grid
with grid spacing $h_x=h_y=h_z=h$ equal along all the axes. 
If the radius of the particles is comparable with $h$ we can assume that
the 3D Navier-Stokes equations take the form
\begin{equation}
\cfrac{D\mathbf{u}}{Dt}=-\cfrac{\nabla p}{\rho_{0}} + \nu\Delta \mathbf{u} + 
\sum_{i = 1}^{N} \frac{\mathbf{F}_i}{\rho_0 h^3} \Phi(\mathbf{x}-\mathbf{x}_i),
\label{NSdelta}
\end{equation}
where $\rho_{0}$ is the fluid density, $\nu$ the kinematic
viscosity, $\mathbf{F}_i$ is the force applied on the fluid by the sphere in
$\mathbf{x}_i$. As typical with immersed boundary methods, the forces are regularized by spreading their effects on 
the nearby grid points with the function $\Phi(\mathbf{x})$, which has the
following properties: $\Phi(\mathbf{x})\geq 0$; $\Phi(\mathbf{x})=0$ for
$|\mathbf{x}|>n h$, with $n$ not necessarily integer, i.e. it has support over
a finite stencil surrounding the particle; normalization, i.e. 
$\sum_{\mathbf{x} \in grid} \Phi(\mathbf{x}-\mathbf{x}_s)=1$ (sum over the points $\mathbf{x}$ 
of the numerical discretized domain)
\cite{peskin1977numerical,peskin1980modeling}.
Following \cite{roma1999adaptive} we use
\begin{equation}
\Phi(\mathbf{x})=
\begin{cases}
\frac{1}{3} \left(1+\sqrt{-3|\mathbf{x}|^2+1}\, \right), & |\mathbf{x}|\le0.5 h \\[8pt]
\frac{1}{6} \left(5-3|\mathbf{x}|-\sqrt{-3(1-|\mathbf{x}|)^2+1}\, \right), & 0.5 h \le|\mathbf{x}|\le1.5 h \\[8pt]
0, & \text{otherwise}.
\end{cases}
\label{fhi}
\end{equation} 

The Lagrangian problem associated with the motion of the swimmer requires the
knowledge of the fluid velocity $\mathbf{u}(\mathbf{x}_i)$ at the position of each bead,
which is defined as a weighted average of the fluid velocity surrounding the 
bead 
\begin{equation}
\mathbf{u}(\mathbf{x}_i)=\sum_{\mathbf{x} \in grid} \mathbf{u}(\mathbf{x}) \Phi(\mathbf{x}-\mathbf{x}_i).
\end{equation}

For what concerns the forces, as discussed above we consider two kinds of 
beads, for the flagellum and the body respectively, whose interactions with the fluid are treated differently.
A flagellum bead is characterized by a constant propulsion force contributing an acceleration $\mathbf{f}_p$ applied on the bead along the swimming direction. An 
equal and opposite force is applied on the fluid to guarantee conservation of momentum.  
A body bead is instead part of a material
boundary along which the natural no-slip condition applies. 
In line with the IB strategy, each body
bead is subjected to the acceleration $\mathbf{f}_{IB}=\beta(\mathbf{u}(\mathbf{x}_i)-\mathbf{v}_i)$,
where $\mathbf{v}_i$ is the velocity of the $i-$th bead and $\beta$ is a large, positive
numerical parameter. Also in this case a 
 force of opposite sign is applied to the fluid. Such IB forces lead to the
reciprocal relaxation, with a characteristic time $\beta^{-1}$ of bead and fluid velocities to the same values, thus
enforcing the no-slip condition.  Clearly $\beta$ affects the relative error on the implementation of the no-slip condition. If the swimmer
moves with a constant swimming velocity ${\mathrm v}_s$ in a still fluid the IB forces are the equivalent of the viscous drag forces so one must have 
 and $f_{IB}\simeq {\mathrm v}_s/\tau_S$, with $\tau_S$ an effective Stokes time of a bead which can be estimated from the parameters obtained with the fitting procedure described below.
This implies that $|\mathbf{u}(\mathbf{x}_i)-\mathbf{v}_i|/{\mathrm v}_s\simeq (\tau_S\beta)^{-1}$. 

The resulting equations of motion for a 3-bead swimmer are: 
\begin{equation}
			\begin{cases}
				\dot{\mathbf{v}}_{1}=f_p\mathbf{n}_1 + \lambda_{12} \mathbf{n}_1+g\mathbf{t}_1\\
				\dot{\mathbf{v}}_{2}=- \lambda_{12} \mathbf{n}_1 + \lambda_{23} \mathbf{n}_2-\beta\left(\mathbf{v}_2-\mathbf{u}(\mathbf{x}_2)\right)+g\mathbf{t}_2\\
				\dot{\mathbf{v}}_{3}= - \lambda_{23} \mathbf{n}_2 - \beta\left(\mathbf{v}_3-\mathbf{u}(\mathbf{x}_3)\right)+g\mathbf{t}_3		
				\label{3bead}
			\end{cases}	
\end{equation}	
In these equations $\lambda_{ij}$ denotes the Lagrange multiplier
associated with the inextensibility of the rod connecting beads $i$ and $j$,
$\mathbf{n}_1$ and $\mathbf{n}_2$ are unit vectors parallel to the rods and $g\mathbf{t}_i$
 are stiff elastic forces acting normal to the rods and
implementing the constraint of fixed angle $\phi$ (see Fig.~\ref{fig1}).
The single terms are discussed in details in \ref{inextensibility}.
The evolution of the Eulerian velocity field is realized by means of a standard,
fully de-aliased, pseudo-spectral code \cite{canutospectral,boyd2001chebyshev}.
Although both the model and its integration are fully three-dimensional, in the following, for the sake of simplicity in visualizing the results, we will restrict the dynamics to the 
$(x,y)$ plane by a suitable choice of the initial conditions for the swimmers.

The rhs of both \eqref{NSdelta} and \eqref{3bead} have the dimension of forces
per unit mass. As detailed above, each force $\mathbf{F}_i$ in the last term of
\eqref{NSdelta} is due to conservation of momentum and is the opposite of
forces acting on the beads and causing the propulsion acceleration or  the
relaxation to fluid velocity.
If we denote with $\mathbf{f}_i$ one of those Lagrangian accelerations in \eqref{3bead}, one must have $\mathbf{F}_i=-m \mathbf{f}_i$ where $m$ is the bead's mass.
For a spherical bead we can write
\begin{equation}
\cfrac{\mathbf{F}_i}{\rho_{0}h^3} = 
-\mathbf{f}_i \cfrac{\rho}{\rho_{0}h^3}\cfrac{4}{3}\pi R^3 \equiv -\mathbf{f}_i c,
\label{eq6}
\end{equation}
where $\rho$ and $R$ are the bead's density and radius, respectively and
$c=\frac{4}{3}\pi \frac{\rho}{\rho_0} \left(\frac{R}{h}\right)^3$
 determines the relative intensity of Lagrangian acceleration and feed-back on
the fluid. For simplicity, in the following, we consider  neutrally buoyant swimmers only ($\rho=\rho_0$).

\begin{figure}[ht!]
	\begin{subfigure}{.4\textwidth}
		\centering
		\includegraphics[trim={0 -2.4cm 0 0},width=0.8\textwidth]{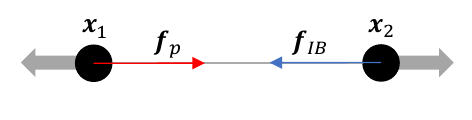}
		\put(-150,10){(a)}
	\end{subfigure}
	\begin{subfigure}{.5\textwidth}
		\centering
		\includegraphics[trim={0 2cm 0 2cm},width=1.2\textwidth]{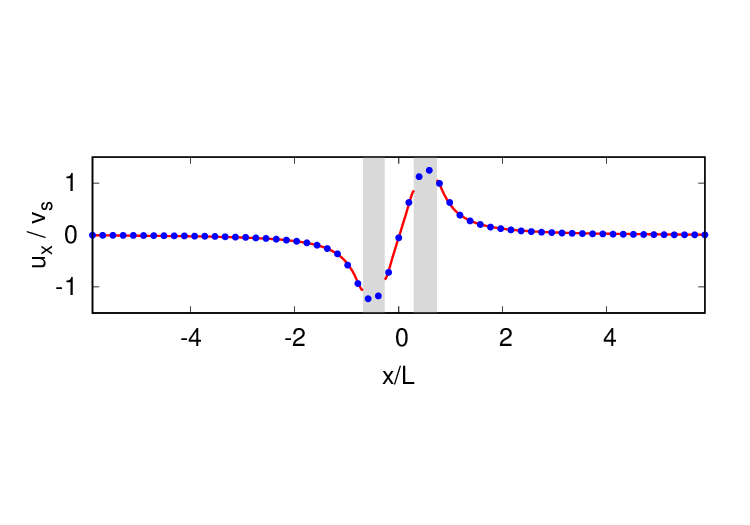}
		\put(-260,10){(b)}
	\end{subfigure}
\caption{(a) Scheme of a two-sphere pusher. The
	gray arrows are the forces exerted by the pusher on the fluid.
(b) x-component of the velocity field along the swimming
direction produced by the two-sphere pusher swimming at velocity ${\mathrm v}_s=0.0065$
with $Re=10^{-2}$. The red line represents the analytical Stokes solution,
blue points are the numerical values computed with $c=5.58$ with 
resolution $64^3$.
}
\label{fit}
\end{figure}

We fix the numerical parameters by considering the simpler case of a pusher composed by two beads
connected by a rigid, inextensible rod (see Fig.~\ref{fit}(a)).  
One of those beads represents the flagellum and one the body and this 
configuration produces two opposite forces on the flow and, therefore,
an approximate force dipole field which decays as $r^{-2}$ in space 
\cite{drescher2010direct}. 
Let $1$ and $2$ be the index of the flagellum and body beads respectively,  
in this case the equations of motion \eqref{3bead} simplify to
\begin{equation}
\begin{cases}
	\dot{\mathbf{v}}_1 = f_p\mathbf{n} + \lambda \mathbf{n}\\
	\dot{\mathbf{v}} = - \lambda \mathbf{n}-\beta(\mathbf{v}_2-\mathbf{u}(\mathbf{x}_2)),
\end{cases}	
\label{syst2ball}
\end{equation}	  
where $\mathbf{n}= (\mathbf{x}_2-\mathbf{x}_1)/|\mathbf{x}_2-\mathbf{x}_1|$ and $\lambda$ is the Lagrange multiplier associated with inextensibility. From this point on, we will use the simplified notation $\mathbf{u}(\mathbf{x}_i)=\mathbf{u}_i$.

The value of $c$ in \eqref{eq6} can be fixed by using the
(approximate) analytical solution of the Stokes flow around the two
spheres. Let $L$ represent the distance between the spheres moving at
velocity ${\mathrm v}_s$.  We consider the swimmer Reynolds number ${\rm
  Re}={\mathrm v}_sL/\nu=10^{-2}$ and we compute the longitudinal component of
the fluid velocity along the axis of the swimmer.  Periodicity of the
domain is taken into account by considering the images in the three
directions. Figure~\ref{fit}(b) shows the
comparison between the analytical (discussed in
  \ref{stokeslets}) and numerical results, which gives the fit
$c\simeq 5.58$, corresponding to $R \simeq 1.1 h$. The analytical
solution in the regions within the effective radius of the beads (the
gray regions in Fig.~\ref{fit}(b)) is excluded
from the comparison since it is singular and unphysical. The numerical
solution, on the other hand, is well behaved also in those regions.
We have tested the consistency of the definition of $c$ by verifying
that it is not affected by the resolution of the grid (up to $256^3$
points) and it is also independent on ${\rm Re}$ when ${\rm Re}
\lesssim 1$.

\section{Numerical results for rectilinear swimmers}
Here we focus on the case of rectilinear swimmers composed by three spheres
connected by two rods of length $L$. In Fig.~\ref{fig1} bead $1$ represents the
flagellum, while beads $2$ and $3$ define the body.  The whole system is
considered rigid and inextensible. Inextensibility is enforced via Lagrange
multipliers while bending rigidity is guaranteed via stiff springs, which is
sufficient to prevent oscillations.  The length of the swimmer is defined as
the distance $L$ between the two beads of the body, thus neglecting the
presence of the flagellum. Therefore, the Reynolds number is defined as in the
case of two-sphere model ${\rm Re}={\mathrm v}_s L/\nu$.

\begin{figure}[h!]
  \begin{subfigure}{0.5\textwidth}
    \centering
    \includegraphics[width=0.9\textwidth]{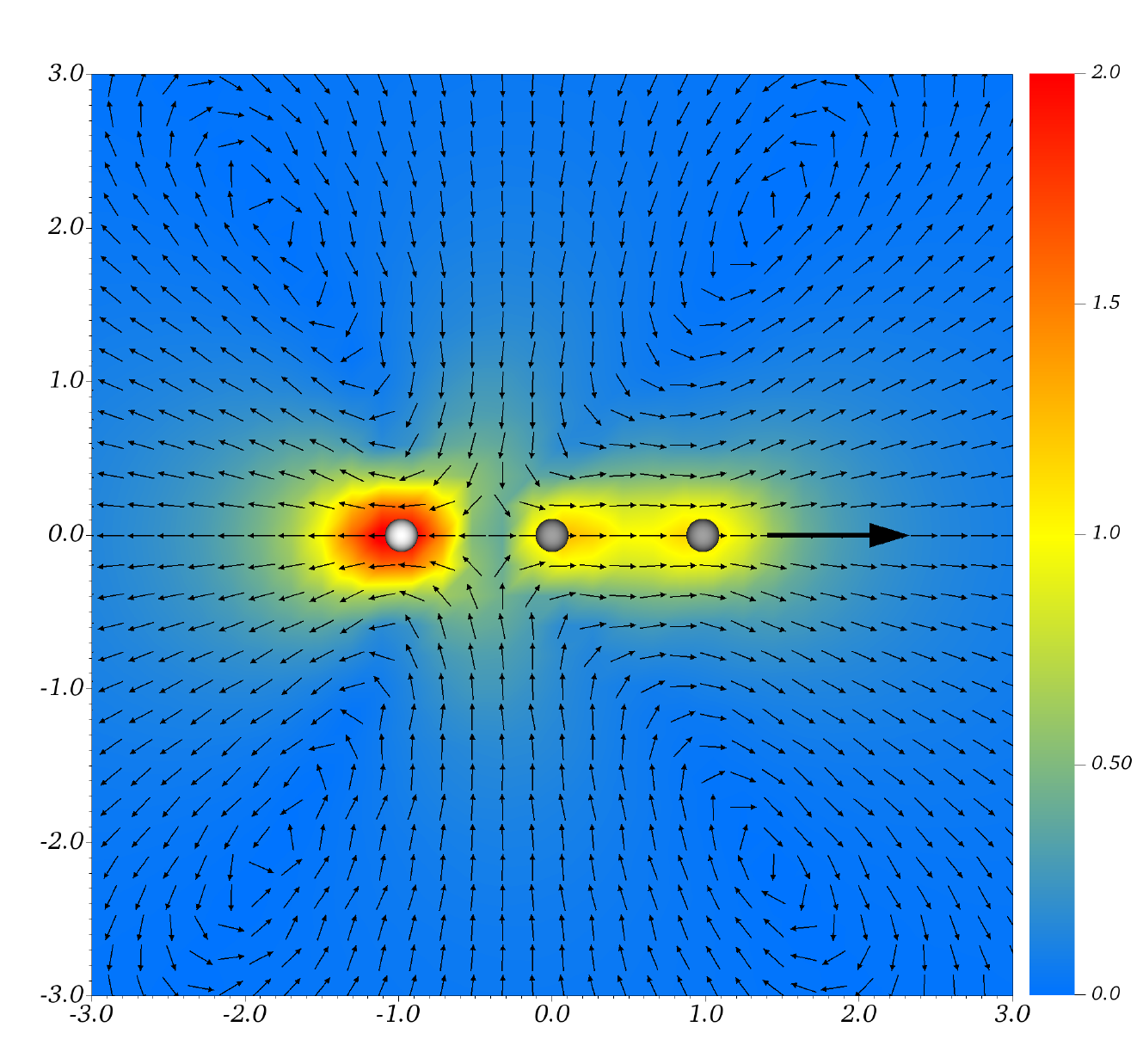}
    \setlength{\unitlength}{1cm}
    \put(-4.3,-0.2){$x/L$} \put(-8.3,
    3.3){\rotatebox{90}{$y/L$}}
    
  \end{subfigure}
  \begin{subfigure}{0.5\textwidth}
    \centering
    \includegraphics[trim={0 0 0 3cm},width = \textwidth]{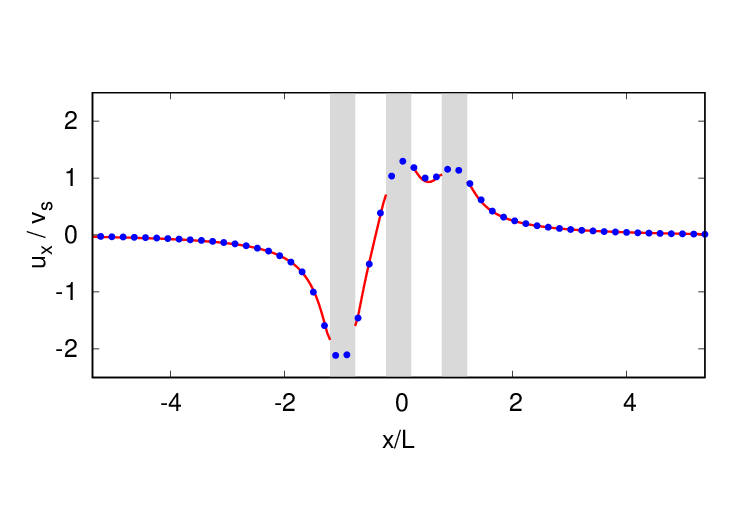}
    \put(-450,-5){(a)}
    \put(-200,-5){(b)}
   
  \end{subfigure}
  \caption{Velocity field surrounding a 3-bead swimmer at the
    stationary state. Velocities are rescaled with the constant
    swimming speed. (a) The color map indicates the amplitude of the
    velocity, while the arrows (not scaled with amplitude) indicates
    the velocity direction. The typical {\em pusher} configuration
    (outwards streamlines along the swimmer axis, inwards in the normal
    direction) can be clearly appreciated.  The swimmer is moving to
    the right. The leftmost white bead represents the flagellum, where
    propulsion is applied. The corresponding reaction force on the
    fluid produces an intense velocity perturbation (red region). The
    black arrow on the right stands for the swimming direction. (b)
    Plot of the x-component of the velocity field along the swimming
    direction of a 3-beads swimmer.  The numerical solution (circles)
    is compared with the approximate analytical solution (continuous
    line, see text). Note that, as expected the fluid field on the
    tail beads (on the left) is comparable with the sum of the
    velocity field produced by the others, as a consequence of
    inextensibility and conservation of momentum.
}
\label{fig3}
\end{figure}

In Fig.~\ref{fig3} we show the results of a numerical simulation of a single
3-bead pusher moving with constant velocity in an otherwise quiescent fluid.
In Fig.~\ref{fig3}(a), a 2D section of the 3D domain containing the swimmer
is shown.  At the stationary state, from Eq.~\eqref{3bead} one must have
$\mathbf{f}_p=-\beta(\mathbf{v}_2-\mathbf{u}_2)-\beta(\mathbf{v}_3-\mathbf{u}_3)$.
In this case the distribution of forces among the three spheres is less trivial
than the completely symmetric case of the 2-sphere model. Figure~\ref{fig3}
shows that the velocity produced by propulsion around the flagellum bead is, in
agreement with the above relation, larger in magnitude than the disturbance
produced by viscous drag around each of the body beads and it is comparable
with the sum of the velocity field produced by the others, according to the
rigidity condition and to the conservation of momentum. Further details on the
implementation of  inextensibility and rigidity can be found in
\ref{inextensibility}.

As detailed in \ref{Appendice:jeff}, in the absence of propulsion, i.e. when the flagellum does not  play any role, the body of the swimmer behaves as an infinitely thin rod and its dynamics is ruled by Jeffery's equation \cite{jeffery1922motion}. Indeed, the equation for the swimming direction $\mathbf{n}$ 
\eqref{jeffery} can be written as 
\begin{equation}
\dot{\mathbf{n}}=\frac{1}{2}\mathbf{\omega} \times {\mathbf{n}} + \Lambda[\mathbb{S}{\mathbf{n}}-({\mathbf{n}}\mathbb{S}{\mathbf{n}}){\mathbf{n}}] 
\label{eq:jeff}
\end{equation}
where $\Lambda$ is the shape parameter ($\Lambda=0$ for spheres and $\Lambda=1$
for infinitely thin rods) and $\mathbf{\omega}=\nabla\times \mathbf{u}$ is the
vorticity. Using a 2D reference system in which the swimming angle is measured
from the horizontal direction ($n_x=\cos \theta$, $n_y=\sin \theta$) one can
write
\begin{equation}
\dot{\theta}=-\frac{\sigma}{2}[1+\Lambda(1-2\cos^2\theta)]\,,
\label{dottheta}
\end{equation} 
where $\sigma$ is the fluid velocity shear rate. If $\Lambda=1$,
$\theta=0$ is a marginally stable fixed point. 
The solution for a rod that starts perpendicular to the shear direction 
is given by
\begin{equation}
\begin{cases}
\theta(t) =\mathrm{arccot}(\sigma t)\\
\theta(0) =\pi/2.
\label{jeffsol}
\end{cases}
\end{equation}   
We expect Eq.~(\ref{jeffsol}) to describe the motion of a three-bead swimmer in
a linear shear when the propulsion is switched off. We stress that the 
flagellum bead can be completely disregarded in this regime.

\begin{figure}[h!]
\centering
\includegraphics[width = 0.8\textwidth]{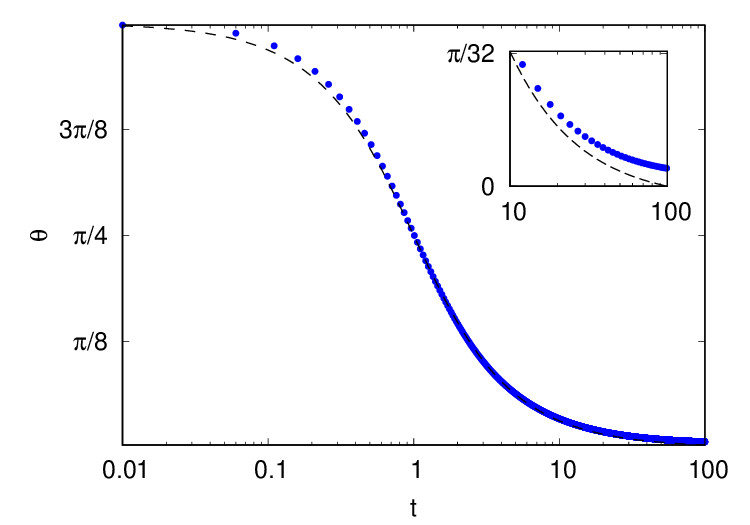}
\caption{Comparison between the solution \eqref{jeffsol}
 of Jeffery's equation for a rod  (formally, an ellipsoid with $\Lambda=1$, dashed line) and numerical data from
the 3-beads model (blue dots) without propulsion in a Kolmogorov flow.
Numerical simulations are done at resolution $N=64^3$ grid points in a 
cube of size $2 \pi$ for a swimmers of length $L=0.5$ with initial 
orientation $\theta=\pi/2$. Inset: zoom of the long time behavior, see text for a discussion.
}
\label{jeff}
\end{figure}

Figure~\ref{jeff} shows the time evolution of the orientation of a
non-motile swimmer compared with the analytical solution
\eqref{jeffsol}.  The numerical solution is obtained by the
integration of a three-bead swimmer with ${\bf f}_p=0$ placed in the
inflection point at $z=3\pi/2$ of a Kolmogorov flow of period $2\pi$ with velocity
$\mathbf{u}=(\cos(z),0,0)$ corresponding to $\sigma=1$
  \cite{borgnino2022alignment,meshalkin1961investigation}.  The
numerical result in Fig.~\ref{jeff} is compared with
the analytical expression \eqref{jeffsol} valid for an ideal rod-like
particle.  The deviations can be quantified by observing that the time
it takes for the numerical swimmer to reach $0.1 \rm{rad}$ is only
$10\%$ larger than the theoretical prediction. Such a small difference
should be all but irrelevant when time dependent flows are
considered. We conclude that for our model swimmer, a linear shear in
a creeping flow regime gives rise to a dynamics that can be described
by Jeffery's equation \eqref{eq:jeff} with $ \Lambda=1$

We now consider the interaction between two swimmers and the resulting
trajectories. We only consider the effects of
  hydrodynamic interactions without any additional repulsive potential
  to account for steric interactions. The latter can be anyway added
to the model in straightforward ways.  We remark that
in our model swimmer there are no physical rods
connecting the beads. Therefore, in principle,
  swimmers can overlap with crossing trajectories.
Nevertheless, we find that, if the beads are not too
far apart, swimmers feel each others as effective {\it continuous}
bodies, thanks to the flow produced in their motion and overlaps are
observed only in very special conditions.

\begin{figure}[h!]
\centering
\includegraphics[width=0.45\textwidth]{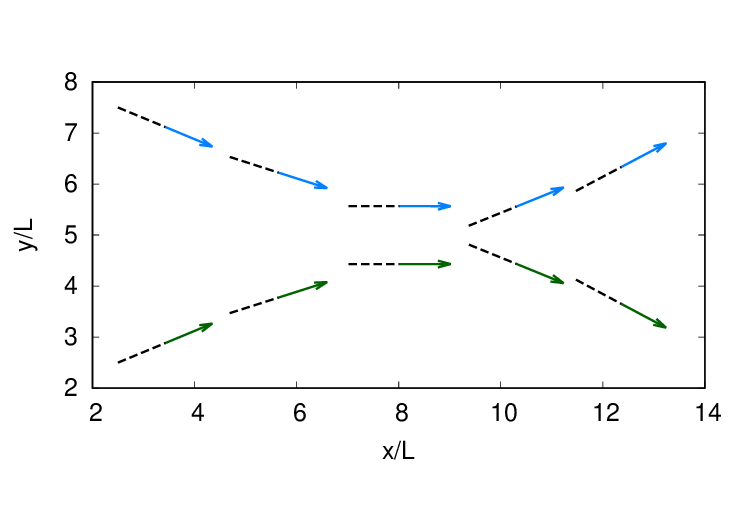}
\includegraphics[width=0.45\textwidth]{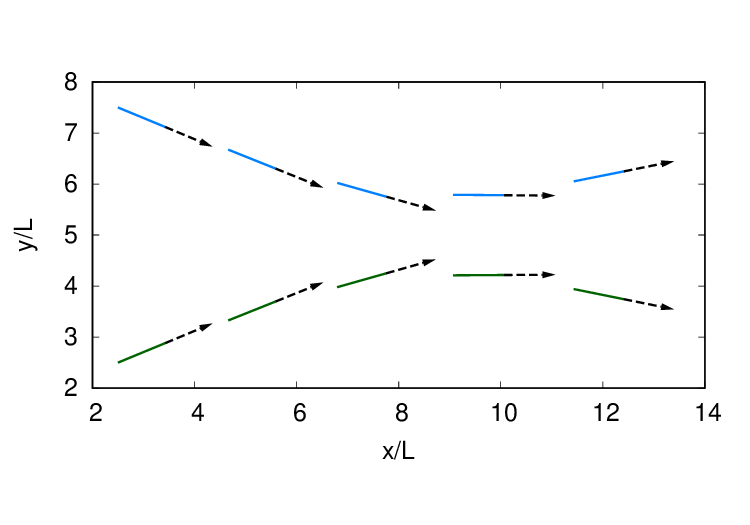}
\put(-12cm,0cm){\footnotesize(a)}
\put(-4cm,0cm){\footnotesize(b)}
\caption{
Collision of two identical swimmers starting with a relative
angle of $\pi/4$. The hydrodynamic interaction between by the body's beads
allows the swimmers to scatter without touching each other. (a) Pusher dynamics. The subsequent positions of the swimmers are plotted, left to right, at regular time intervals. The solid
arrows indicate the inextensible rod connecting the body beads, while the dashed lines
represent the flagella, connecting the rear
body beads to the flagellum beads. Only the body segment is considered as a
rigid boundary, on which the no-slip condition is applied for the fluid. (b) Pullers dynamics. The same time series of pushers' case is shown. The interaction in this case produces a smaller output angle due to the different hydrodynamic interaction between the swimmers.
}
\label{fig5}
\end{figure}

We start by considering the scattering
of two identical swimmers, moving at the same speed, with an incident
angle $\theta_i$. One example is shown in
Fig.~\ref{fig5}(a) with $\theta_i=\pi/4$.  The scattering is a
complex process during which the two swimmers orient temporarily in a
parallel direction and finally emerge with a different output angle
$\theta_o$. In the case of pusher swimmers, the velocity field (see
figure \ref{fig3}(a)) causes the flagella to come closer together,
turning the swimmers and leading first to the alignment of the
swimmers and subsequently to a separation of the
directions. The above described phenomenology is
consistent to what found in \cite{gyrya2010model,furukawa2014activity}
starting from parallel swimmers.  For a pair of
  pullers, the kinematics is qualitatively very similar, except that
  the hydrodynamics which produces it is opposite to that of pushers,
  see Fig.~\ref{fig3}(b). The flagella, which in this case are
the first to interact, tend to repel each other, leading to the same
kinematics of alignment and subsequent divergence of the
trajectories. The exit angle is consequently different in the two
cases, as evident from comparing Figs.~\ref{fig3}(a) and (b).  We
remark that the model does not exclude the possibility of observing
the overlap between swimmers under certain conditions (for example, in
the case of high Reynolds numbers or very large collision angle). The
most common case is a superposition of the flagella. This event is not
per se problematic because in our model the flagella are not affected
by hydrodynamic interactions. We stress that with our method (and at
variance with other approaches \cite{furukawa2014activity} where the
particles have a finite volume which behaves as a second fluid with a
large viscosity) the beads are effectively represented as regularized
point forces whose effective radius is a numerical parameter used to
fit the resulting velocity field. The occasional partial overlap of
the force stencils can therefore cause numerical stiffness, by
introducing large local forces, but is not necessarily physically
inconsistent. The cases in which also the bodies overlap can be
avoided with an effective, short-range repulsion potential. Such
potential can take different forms essentially corresponding to steric
interactions between the beads or between the bodies (through the
definition of an effective shape). We consider here only the effects
of hydrodynamic interactions and remark that no numerical instability
was observed as a result of the overlap of the tails or the bodies in
the case of binary collisions.

\section{Circle swimmer trajectories}
\label{turning}
The 3-beads model allows to control the 
swimming direction in a simple and natural way. Indeed, when the three beads are 
not in a collinear configuration, the drag on the body together with the
propulsion from the flagellum produce a torque that rotates the swimmer. 
In what follows we will present only results about circle pushers, in which the flagellum bead is the trailing one. As discussed above, puller-like circle swimmers can be obtained by reversing the propulsion.
The controlling parameter for the swimmer is the equilibrium angle 
$\phi_0$ between
the flagellum rod and the body rod.
The resulting trajectory of a single swimmer is a circle with a radius
$R_c$ depending on $\phi_0$. Clearly in the limit $\phi_0\to 0$ one recovers the original collinear model, with $R_c \to \infty$.

Once the bending angle is set, the rigidity of the swimmer is guaranteed by an elastic
force that causes the relative position of the two rods to relax to that angle.
This elastic force is implemented in the form of 
internal forces $\mathbf{g}_i$, one for each bead (see~\ref{inextensibility}).
Referring to \eqref{g}, this means that if a perturbation
produces deviations from the equilibrium angle $\phi_0$ these 
are compensated by the
torques due to the internal forces, bringing the system to the wanted
configuration. 
It is worth noting that equation \eqref{g} should produce a harmonic oscillation of the angle
$\phi$ around $\phi_0$. These oscillations are 
damped by viscosity through the no-slip condition on the body beads, thus causing a relaxation to the prescribed angle $\phi_0$.

\begin{figure}[hb!]
	\centering
	\includegraphics[trim={1cm 0 0 0},width=0.32\textwidth]{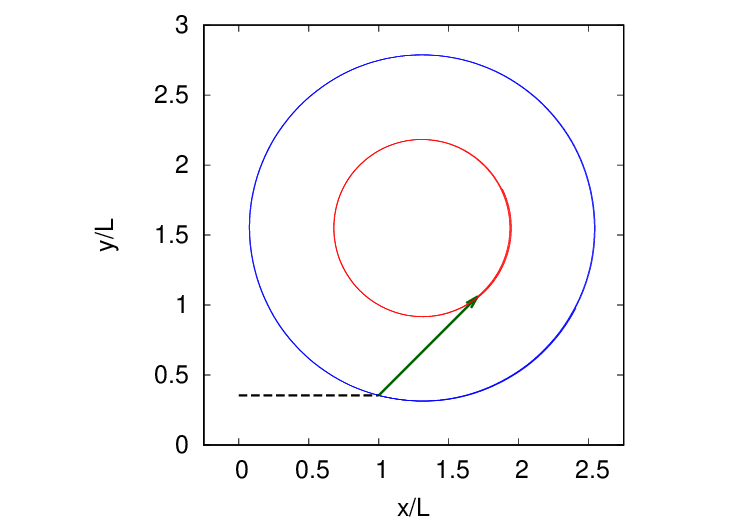}
	\includegraphics[trim={1cm 0 0 0},width=0.32\textwidth]{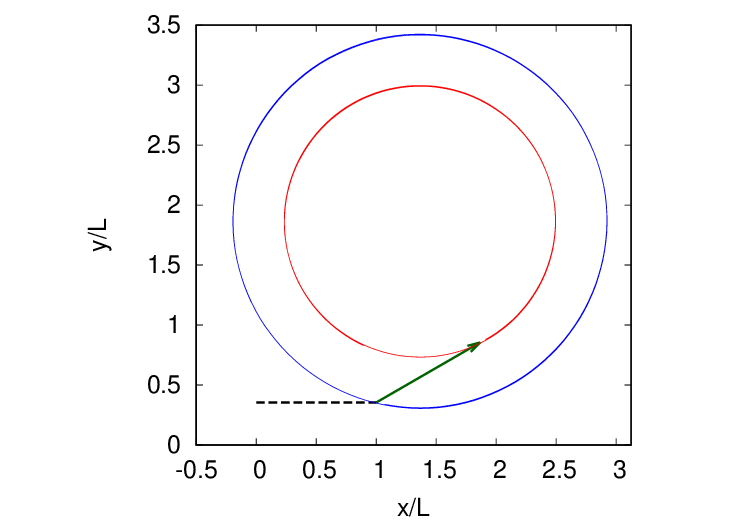}	
	\includegraphics[trim={1cm 0 0 0},width=0.32\textwidth]{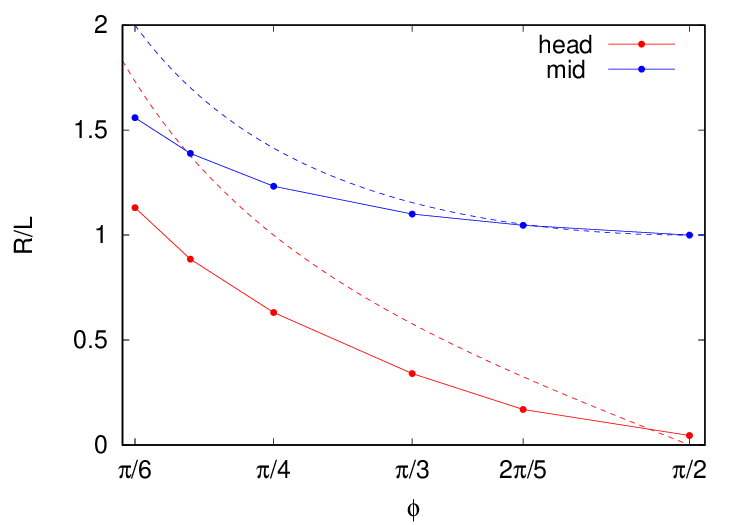}	
	\put(-13.5cm,4.5cm){\footnotesize(a)}
	\put(-8.2cm,4.5cm){\footnotesize(b)}
	\put(-2.7cm,4.5cm){\footnotesize(c)}
	\caption{Trajectories of isolated swimmers with a constant bending
angle. (a,b) Two typical trajectories, with $\phi=\pi/4$ and  $\phi=\pi/6$,
respectively. The red (blue) lines mark the trajectory of the head-bead
(mid-bead). (c) Dependence of the radii of the trajectories of the head and mid beads (solid lines) on the angle $\phi$
between the flagellum and the body. The results obtained with a simple model
(dashed lines, see text) are given for comparison.} 
	\label{girotondo}
\end{figure}

Figure~\ref{girotondo} shows two examples of circular trajectories
produced by circle swimmers with different bending angles, together with the dependence 
of the radius of the trajectory on $\phi$. Observing that the
segments identifying the body and the flagellum are approximately tangent to
the circles described by head and middle beads, respectively, one can
tentatively estimate the radii of those circles as $r_{\rm
	head}\sim L/\tan(\phi)$ and
$r_{\rm mid}\sim L/\sin(\phi)$. The actual radii (see Fig.~\ref{girotondo}(c)) are smaller than the estimate (dashed lines) except for $\phi\sim\pi/2$, in which case the head is almost stationary ($r_{\rm head}\sim 0$) and the flagellum rotates by remaining approximately tangent to the outer circle ($r_{\rm mid}\sim L$).
More complex trajectories can be obtained if we allow the
angle to change in time, and this can be used to control the 
swimming trajectory. We leave the dynamics of active steering for future investigations.

\begin{figure}[h!]
\centering
\includegraphics[trim={2.5cm 0 0 0cm}]{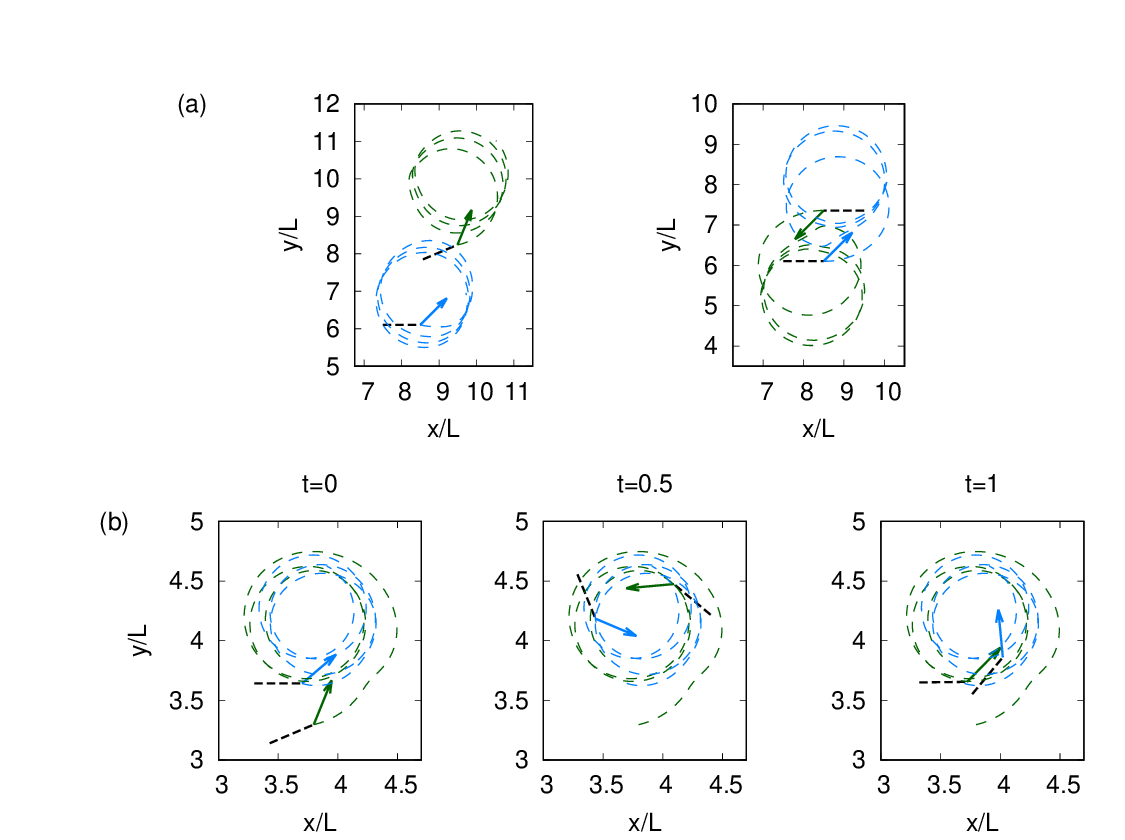}		
\caption{Interaction between a pair of circle swimmers. (a) Two configurations
are shown in which after a transient the two swimmers settle onto essentially
independent trajectories. (b) Dynamics of two swimmers starting initially very
close and nearly parallel. In this case the two trajectories are intertwined
and revolve around a common center. Note how the relative positions of the
swimmers change during their orbits. The position is rescaled with the rod
length between two beads and times are rescaled with the typical time in which
the swimmer covers its length.}
\label{2bentb}
\end{figure}

In analogy with the case of a straight swimmer, we studied the interaction of
two circle swimmers. The bending angle is fixed at $\pi/4$ as shown in Figure
\ref{girotondo}. Two different behaviors were observed, depending on the
relative initial positions of the swimmers. If the initial separation is large
enough, as expected, each swimmer tends to swim on its own curvilinear trajectory without interacting, in some cases after a brief transient characterized by a repulsive interaction. Two examples of this behavior are shown in Fig.~\ref{2bentb}(a), corresponding to different initial conditions. 
If the initial separation is further decreased (Fig.~\ref{2bentb}(b)) the interaction changes
qualitatively. After a more complex initial transient, the two trajectories
intertwine and start revolving around the same center. It is worth noting that the reciprocal positions of the swimmers are not locked along the orbit but change dynamically in a non-trivial way. In the presence of many
swimmers, a random initial configuration can lead one swimmer to decouple from
one neighbor (as in Fig.~ \ref{2bentb}(a)) only to be attracted by another one
into forming a strongly coupled pair (as in Fig.~\ref{2bentb}(b)). This
mechanism could lead to an ordered collective behavior as briefly discussed in the next section.

\section{Collective behavior}
\label{collective}
The numerical method proposed in this work can be easily scaled to a large
number of swimmers to study their interaction and the emergence of 
collective motion. 
As an example we considered $500$ identical pushers initially placed at 
random positions and directions on a $(x,y)$ plane in the 3D domain. In the absence of perturbations in 
the $z$ direction, the motion remains planar, thus confirming the accuracy of the numerical integration.   

\begin{figure}[h!]
	\centering
	\includegraphics[trim={17cm 0 0 0},width=1.2\textwidth]{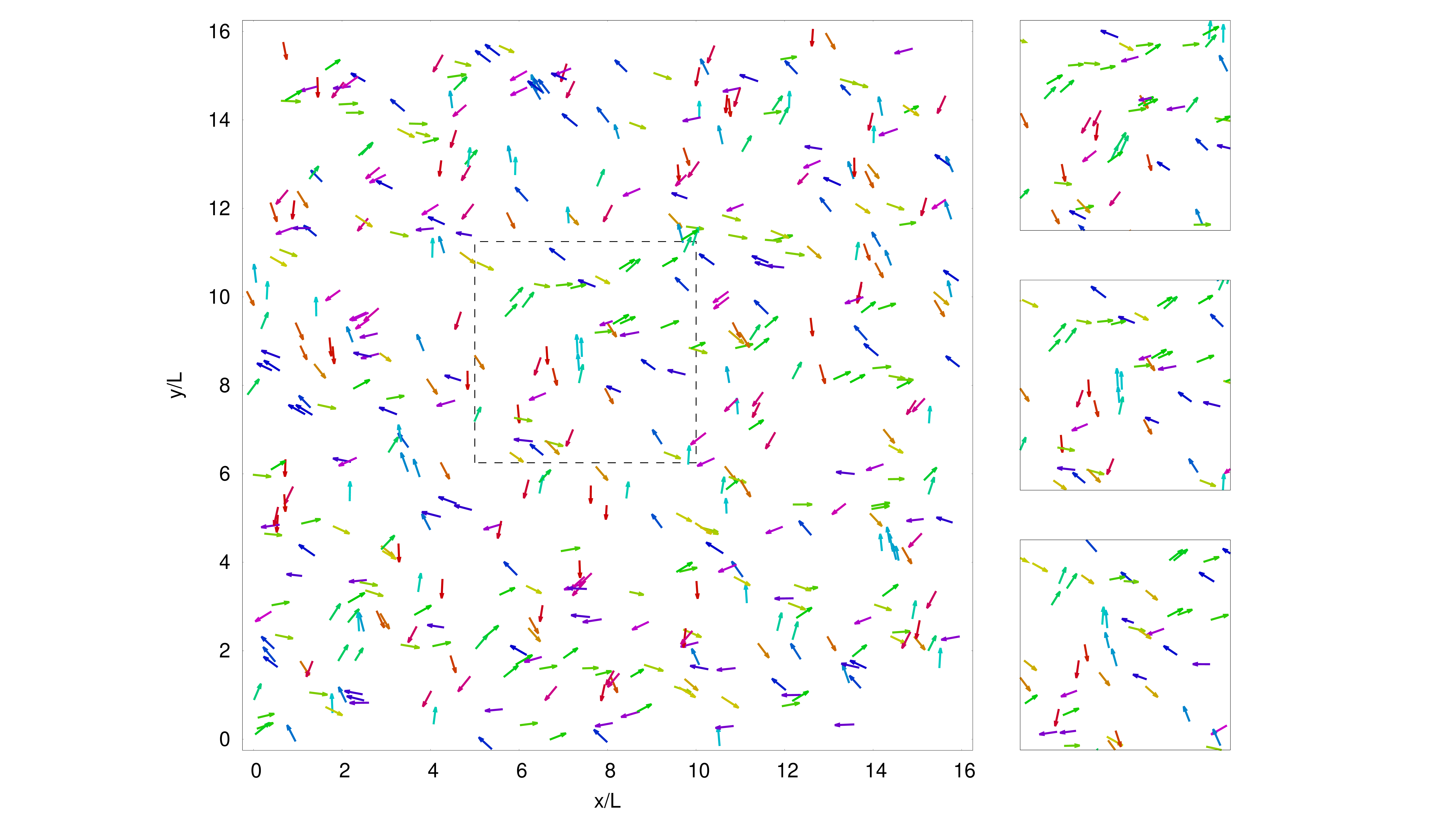}
	\put(-550,0){(a)}
		\put(-200,260){(b)}
	\put(-200,140){(c)}
	\put(-200,10){(d)}
	\caption{Collective behavior of 500 swimmers in a 2D configuration
within a 3D fluid domain. In this figure the flagella are not drawn and each swimmer is colored based on its angle with respect to the x axis, so that parallel swimmers have the same color. The fluid is forced into a chaotic flow by the
motility of the swimmers. The resulting velocity fluctuations induce relatively
large velocity differences between nearby swimmers which occasionally defeat
the repulsive effect of hydrodynamic interactions and cause the bodies to
overlap more frequently. The formation of clusters of schools of swimmers sharing the same swimming direction is highlighted by the coloring scheme. On
the left a 2D snapshot of the whole domain at time $t\simeq23$ (rescaled with
the typical time in which the swimmer covers its length) is shown. (b), (c) and (d): three zoomed snapshots of the dynamics within the dashed square in panel (a). Panel (b) and (c) are taken at a time interval $\Delta t=0.30$ before and after the main panel (corresponding to panels (a) and (c)), respectively. It is here evident that the schools persist several swimmer lengths following the surrounding dynamics.
	} 
	\label{fig8}
\end{figure}

One snapshot of the configuration of the swimmers at late time 
is shown in Fig.~\ref{fig8}. We observe that the 
distribution is not random any more, with local
clusters (or schools) swimming in a parallel direction, similarly to the intermediate
state observed in Fig.~\ref{fig5}.
This configuration is highly dynamical, as different clusters appear and dissolve in time
in a statistically stationary condition (see Fig.~\ref{fig8}(b,c,d)).
In this dense condition confined on 
a plane, the occurrence of overlapping swimmers is not uncommon. 
In a realistic application with the full 3D motion, the overlap would be 
much more occasional as the mean free path of swimmers would be 
much larger. Remarkably, even in the case of Fig.~\ref{fig8} we find that 
the swimmer model does not develop numerical instabilities as a 
consequence of the close encounters. However, when a similar case is studied for
pullers (not shown), the ensuing clustering is much stronger than for pushers \cite{bardfalvy2024collective,skultety2024hydrodynamic} and rapidly leads to numerical
instabilities due to the overlap of a large number of beads, with
their relative force stencils. Clearly in this case a repulsion force must  be
implemented.

\begin{figure}[h!]
	\centering
	\includegraphics[trim={17cm 0 0 0},width=1.2\textwidth]{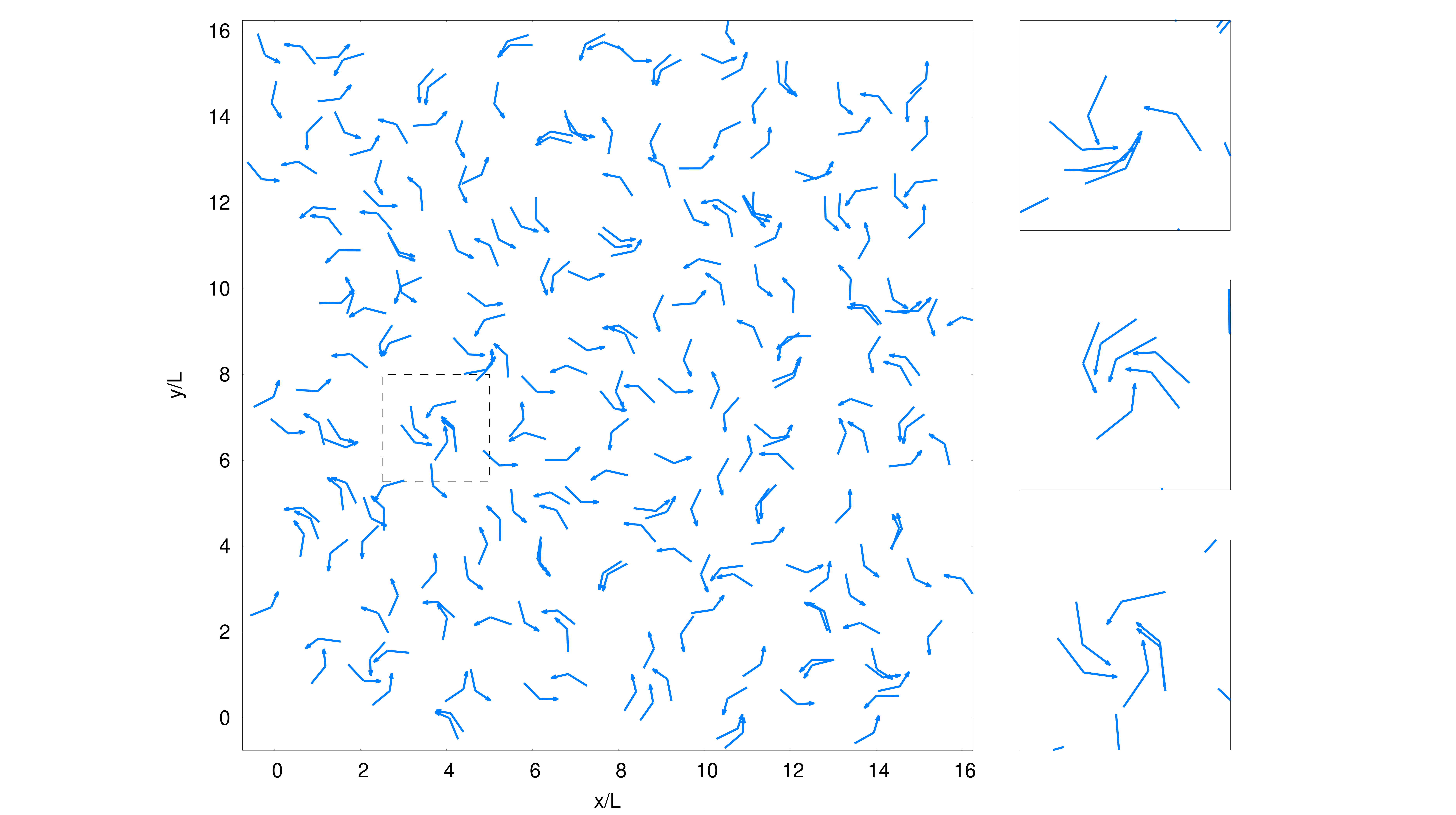}
	\put(-550,0){(a)}
	\put(-200,260){(b)}
	\put(-200,140){(c)}
	\put(-200,10){(d)}
	\caption{Collective behavior of 250 turning swimmers in a 2D
		configuration.  The dynamics is characterized by the formation of groups of
		swimmers that,  thanks to the hydrodynamic interaction, pair up to form groups
		of swimmers that evolve along nearly circular, approximately concentric
		trajectories. These structures are very robust. On the left a 2D snapshot of the whole
		domain at time $t \simeq 16$  (rescaled with the typical time in which the
		swimmer covers its length). On the right three zooms of the dynamics of swimmers. The central plot shares the same time of the snapshot on the left and is preceded in time by the one below it and followed by the one above it, both with a time interval of $\Delta t = 0.30
		$. Indeed these structures are very robust once formed.}
	\label{fig9}
\end{figure}


The discussion in the previous section, as well as previous literature
\cite{kaiser2013vortex, yang2014self}, suggest that circle swimmers can present
interesting collective dynamics. Also in this case we show here only results
regarding circle pushers, since pullers tend to  undergo strong clustering
that requires the implementation of steric interactions. We
considered the case of $250$ circle swimmers (Fig.~\ref{fig9}). The collective
dynamics in this case is characterized by a transient in which swimmers with an
initial condition
similar to those observed in Fig.~\ref{2bentb}(a) tend to move apart until they
intersect other trajectories with which to form a collective circular
trajectory, as shown in Fig.~\ref{2bentb}(b). An example of the resulting collective motion is shown in the side panels of Fig.~\ref{fig9}. Once swimmers achieve this
coupled configuration, the dynamics become rather complex because each orbit is
traveled at different and non-constant speeds. Starting from a configuration
where all swimmers are closely packed (Fig.~\ref{fig9}(b)), the flow generated by each pusher accelerates the
nearby swimmers, causing a fast rotation and a progressive separation
(Fig.~\ref{fig9}(c,d)) of the swimmers along their collective orbit. At later
times a packed configuration forms again.
 This behavior repeats and allows
the formation of these structures on the scale of the swimmer. Preliminary observations suggest that, once formed, these structures tend to persist and produce a global configuration characterized by many swimmer vortices (Fig.~\ref{fig9}(a)). Quantitative assessment of the persistence of the collective structures as well as their statistical correlations is needed, in order to fully characterize this system, and will be the subject of future work.

\section{Conclusions}
In this paper we have proposed and analyzed a numerical model based on immersed
boundary methods of a minimal swimmer, whose body is modeled by two beads and
flagellum represented by a single bead. The model can be used for both pushers
and pullers.   The choice of two beads, with no slip conditions, for the body
make the swimmer to feel the gradient of the velocity field allowing it to be
rotated by the flow. In particular, We showed that, when the propulsion is
switched off, the swimmer moves approximately according to Jeffery's equations
for a thin rod. When the three beads are collinear the model swimmer display
straight swimming, while by maintaining an angle between body and flagellum it
swims in circles. We analyzed the close encounters between both straight and
circle swimmers showing how hydrodynamic interactions mediated by the solvent
fluid make the two swimmers to scatter. Then we scaled up the system by
considering many either straight or circle swimmers and showed that the active
suspension can give rise to non trivial collective motions. For straight
swimmers,  local alignment can be observed in the presence of the sole
hydrodynamic interactions leading to dynamic schools of swimmers swimming in the
same directions. Remarkably, interactions between co-rotating circle swimmers
lead to the formation of approximately ordered vortices of swimmers, moving on
approximately circular trajectories. These are clearly preliminary results and
the collective dynamics of the model should be studied more extensively, also
in light of previous results on circle swimmers
\cite{yang2014self,kaiser2013vortex}. In particular, it will be interesting to
assess whether and to what extent the structures observed with other model
swimmers are model-independent and how the collective dynamics changes with steric interactions. Another interesting direction of investigation is to allow the
swimmers to change dynamically their geometry, this can be used to control the
swimming direction internally so to the swimmer can steer and direct its motion
in a desired direction. Eventually, this can be supplemented by artificial
intelligence, e.g. via reinforcement learning \cite{sutton2018reinforcement} so to allow the swimmers to
accomplish some single (e.g. reach a target or control dispersion
\cite{borra2022reinforcement,calascibetta2023taming}) or collective
goal \cite{durve2020learning} (e.g. swimming in schools). These features can be
useful to microrobots design in biomedical applications, to model animal
interactions etc.

\section*{Acknowledgments}
We thank A. Mazzino and M. Cavaiola for enlightening discussions on the 
IBM.
We acknowledge HPC CINECA for computing resources within the INFN-CINECA Grant 
INFN24-FieldTurb.

\bibliographystyle{RS}
\bibliography{biblio}

\appendix

\section{Stokeslets}
\label{stokeslets}
Let's start by considering a single sphere with a constant speed
$\mathbf{v}$. The velocity field produced, in the very low ${\rm Re}$ regime, is
described by the Stokes equation and leads to the following solution \cite{landau2013fluid}:		
		 \begin{equation}
		 	\mathbf{u}=\frac{3}{4}R \frac{\mathbf{v}+\hat{\mathbf{r}}(\mathbf{v}\cdot\hat{\mathbf{r}})}{r} + \frac{1}{4} R^{3} \frac{\mathbf{v}-3\hat{\mathbf{r}}(\mathbf{v}\cdot\hat{\mathbf{r}})}{r^{3}}
		 	\label{stokeslet}
		 \end{equation}		
		 where $\hat{\mathbf{r}}$ is a unit vector pointing from the center of the sphere (origin) to a point in space, $r$ is the distance with respect to the origin and $R$ is the radius of the sphere. The previous equation could be rewritten as:		
		 \begin{align}
		 	u_{\alpha}&=\frac{3}{4}\frac{R}{r}[{\mathrm v}_{\alpha}+\hat{r}_{\alpha}({\mathrm v}_{\beta}\hat{r}_{\beta})]+\frac{1}{4}\frac{R^{3}}{r^{3}}[{\mathrm v}_{\alpha}-3\hat{r}_{\alpha}({\mathrm v}_{\beta}\hat{r}_{\beta})] \nonumber \\
		 	&=\frac{1}{4}(\frac{3R}{r}+\frac{R^{3}}{r^{3}}){\mathrm v}_{\alpha} + \frac{3}{4} \hat{r}_{\alpha}({\mathrm v}_{\beta}\hat{r}_{\beta})[\frac{R}{r}-\frac{R^{3}}{r^{3}}] 
		 	\label{fluidfield}
		 \end{align}			
		 where the Greek subscript stands for spatial components. On the surface of the sphere $r=R$ the no slip condition $u_{\alpha}={\mathrm v}_{\alpha}$ is enforced. 
The fluid field around a dumbbell swimmer is approximately given by
the superposition of two solutions having the same form of \eqref{fluidfield}.
This approximation clearly breaks down on the surface of the beads because it
violates the no-slip condition, but this is not relevant to our numerical model
because the beads have only an effective radius and their surface is not
resolved. Carrying on with this approximation, we denote by $v_i^{*}$ the speed
of the $i-$th sphere if it were isolated. Taking into account the disturbance
induced by the other bead, one gets a linear relation between these
speeds and the ones resulting from hydrodynamic interaction, formally  
		 \begin{equation}
		 	\begin{cases}
		 		{\mathrm v}_1={\mathrm v}_1^*+{\mathrm v}_2^*S\\
		 		{\mathrm v}_2={\mathrm v}_2^*+{\mathrm v}_1^*S,
		 		\label{effective_speed}
		 	\end{cases}
		 \end{equation}
		 where 1 and 2 are the indices of the flagellum and body beads
respectively and $S$ is a geometric factor which can be computed from
\eqref{fluidfield}. $S$ appears in a symmetric way in both equations because the
beads are identical. Using $\mathrm v_\alpha$ known from the numerical computation, equations \eqref{effective_speed} can be inverted obtain
the unknown
velocities $\mathrm v^*$, which can then be plugged into \eqref{fluidfield} to compute the disturbance field. Thus this two Stokes solutions  are superposed and compared with the numerical velocity field in order to fit the
effective radius $R$ of each sphere. 

\section{Inextensibility and rigidity}
\label{inextensibility}
Here we detail how inextensibility and rigidity are imposed and used to fix the
model parameters.  Consider the 2-beads model discussed in the introduction.
The inextensibility condition is:		
\begin{equation}
	|\mathbf{x}_2-\mathbf{x}_1| = \text{const} \quad \Rightarrow \quad \frac{d}{dt}|\mathbf{x}_2-\mathbf{x}_1| ^{2}=0
	\label{d|x1-x2|}
\end{equation}
from which expanding the square $\mathbf{x}_2-\mathbf{x}_1$ and considering further derivation we obtain a condition on accelerations  		
\begin{equation}
	(\dot{\mathbf{v}}_2-\dot{\mathbf{v}}_1)\cdot \mathbf{n} = - \frac{|\mathbf{v}_2-\mathbf{v}_1|^{2}}{|\mathbf{x}_2-\mathbf{x}_1|}.
	\label{cond}
\end{equation}
In the last equation we have introduced the unit vector $\mathbf{n}=(\mathbf{x}_2-\mathbf{x}_1)/|\mathbf{x}_2-\mathbf{x}_1|$.
Equation~\eqref{cond} is trivial for the 1D case, where a zero relative acceleration leads to a zero relative velocity difference. 

From the \eqref{cond} and \eqref{syst2ball} we get:		
\begin{equation}
	\lambda=\frac{1}{2} \left[	\frac{|\mathbf{v}_2-\mathbf{v}_1|^{2}}{|\mathbf{x}_2-\mathbf{x}_1|}-f-\beta [\mathbf{v}_2-\mathbf{u}_2]\cdot \mathbf{n}	\right]
	\label{lambda}
\end{equation}
that is the module of the tension. From \eqref{syst2ball} it is easy to obtain
the stationary state $f\mathbf{n} = \beta(\mathbf{v}_2-\mathbf{u}_2)$ when
$\dot{\mathbf{v}}_1=\dot{\mathbf{v}}_2=0$. Numerical simulations of the model here
introduced, with the constrain expressed by \eqref{lambda}, were carried out on
a dumbbell $0.5h$ long (where $h$ is the grid step) in a 2D Kolmogorov flow of
period $2\pi$ with velocity $\mathbf{u}=(\cos(z),0,0)$ and we observed a maximum relative deviation of the length of each rod of order $10^{-6}$, which validates the model.

The model can be easily extended to the 3-beads swimmer. 
We define two unit vectors $\mathbf{n}_1$ e $\mathbf{n}_2$ that point 
respectively from the tail to the central bead and from the central to 
the head bead. This model introduces a new degree of freedom that is
the angle $\phi$ between the two unit vectors (see Fig.~\ref{fig1}(c) in the main text). To maintain a rigid shape we need this angle to relax to a fixed value $\phi_0$.
We define the unit vectors $\mathbf{t}_{1}$ and $\mathbf{t}_{3}$
 perpendicular to $\mathbf{n}_{1}$ and $\mathbf{n}_{2}$,
respectively, such that they lie in the plane defined by the swimmer, in formulae:

		\begin{align}
			\mathbf{t}_1&=\frac{\mathbf{n}_2-\cos\phi\mathbf{n}_1}{|\mathbf{n}_2-\cos\phi\mathbf{n}_1|} \\
			\mathbf{t}_3&=\frac{-\mathbf{n}_1+\cos\phi\mathbf{n}_2}{|-\mathbf{n}_1+\cos\phi\mathbf{n}_2|}  
		\end{align}
We introduce $\mathbf{g}_1$ along $\mathbf{t}_1$, $\mathbf{g}_3$ along $\mathbf{t}_3$, and
$\mathbf{g}_2$ such that $\mathbf{g}_2=-(\mathbf{g}_1+\mathbf{g}_3)=-g(\mathbf{t}_1+\mathbf{t}_3)$. In
the last equality we suppose that $|\mathbf{g}_1|=|\mathbf{g}_3|=g$. At each time step
$t$ we compute $g$ as
\begin{equation}
g=-a(\phi-\phi_0), 
\label{g}
\end{equation}   
where $a$ is a constant setting the stiffness of the spring which keeps $\phi$ close to $\phi_0$. The equations of the dynamics for a generic angle $\phi$ are:
\begin{equation}
			\begin{cases}
				\dot{\mathbf{v}}_{1}=f\mathbf{n}_1 + \lambda_{12} \mathbf{n}_1+g\mathbf{t}_1\\
				\dot{\mathbf{v}}_{2}=- \lambda_{12} \mathbf{n}_1 + \lambda_{23} \mathbf{n}_2-\beta(\mathbf{v}_2-\mathbf{u}_2)+g\mathbf{t}_2\\
				\dot{\mathbf{v}}_{3}= - \lambda_{23} \mathbf{n}_2 - \beta(\mathbf{v}_3-\mathbf{u}_3)+g\mathbf{t}_3		
				\label{3spsyst}
			\end{cases}	
\end{equation}	  
where $\lambda_{12}$ and $ \lambda_{23}$ are the tension forces that guarantee the inextensibility.
The condition \eqref{cond} is now applied on each rod and we obtain:
		
		\begin{align*}
			\lambda_{23}= &\, \Biggr [ -\frac{|\mathbf{v}_2-\mathbf{v}_1|^{2}}{|\mathbf{x}_2-\mathbf{x}_1|} -\frac{2}{\cos\phi} \frac{|\mathbf{v}_3-\mathbf{v}_2|^{2}}{|\mathbf{x}_3-\mathbf{x}_2|} + f - \beta(\mathbf{v}_2-\mathbf{u}_2)\cdot\left(\frac{2\mathbf{n}_2}{\cos\phi}-\mathbf{n}_1\right)+\\
			&\frac{2\beta}{\cos\phi}(\mathbf{v}_3-\mathbf{u}_3)\cdot\mathbf{n}_2-g_{2}\mathbf{t}_2\cdot\left(\frac{-2\mathbf{n}_2}{\cos\phi}+\mathbf{n}_1\right) \Biggr ] \cfrac{1}{\cos\phi-\frac{4}{\cos\phi}}\\
			\lambda_{12}= &\, \Biggr [ -\frac{|\mathbf{v}_3-\mathbf{v}_2|^{2}}{|\mathbf{x}_3-\mathbf{x}_2|} -\frac{2}{\cos\phi} \frac{|\mathbf{v}_2-\mathbf{v}_1|^{2}}{|\mathbf{x}_2-\mathbf{x}_1|} + \frac{2f}{\cos\phi} + \beta(\mathbf{v}_2-\mathbf{u}_2)\cdot\left(\frac{2\mathbf{n}_1}{\cos\phi}-\mathbf{n}_2\right)+\\
			&\beta(\mathbf{v}_3-\mathbf{u}_3)\cdot\mathbf{n}_2-g_{2}\mathbf{t}_2\cdot\left(\frac{2\mathbf{n}_1}{\cos\phi}-\mathbf{n}_2\right) \Biggr ] \cfrac{1}{\cos\phi-\frac{4}{\cos\phi}}
		\end{align*}
  
\section{A (short) dumbbell is a Jeffery particle with $\Lambda=1$}
\label{Appendice:jeff}
In this appendix we show that the dynamics of a short dumbbell is well described by Jeffery's equation for an infinitely thin rod, see also
the discussion in Sec. III and in particular Fig.~\ref{jeff}.
Consider a dumbbell with fixed length $L$ and particles $\mathbf{x}_1$ and $\mathbf{x}_2$. Assume the dynamics is Stokesian with relaxation time $\tau$. The equations of motion are
\begin{equation}
	\left\{
	\begin{array}{l}
		\dot{\mathbf{v}}_1=-\dfrac{\mathbf{v}_1-\mathbf{u}_1}{\tau}-\lambda\mathbf{n} \\[10pt]
		\dot{\mathbf{v}}_2=-\dfrac{\mathbf{v}_2-\mathbf{u}_2}{\tau}+\lambda\mathbf{n}
	\end{array}
	\right.
	\label{eq:1}
\end{equation}
where $\mathbf{u}_i=\mathbf{u}(\mathbf{x}_i)$ is the fluid's velocity at the $i$-th particle
and $\mathbf{n}=\mathbf{L}/L$, with $\mathbf{L}=\mathbf{x}_2-\mathbf{x}_1$. The modulus of the
rod's tension $\lambda$ is obtained by imposing inextensibility
$dL^2/dt=0$ (see \eqref{d|x1-x2|}). Further derivation to obtain a condition on accelerations gives
\begin{equation}
	\ddot{\mathbf{L}}\cdot\mathbf{L}+|\dot{\mathbf{L}}|^2=0,
	\label{eq:2}
\end{equation}
where $\dot{\mathbf{L}}=\mathbf{v}_2-\mathbf{v}_1$ and
$\ddot{\mathbf{L}}=\dot{\mathbf{v}}_2-\dot{\mathbf{v}}_1$. Defining 
$\mathbf{w}_i=\mathbf{v}_i-\mathbf{u}_i$, we get from (\ref{eq:1})
\begin{equation}
	\ddot{\mathbf{L}}=-\frac{\mathbf{w}_2-\mathbf{w}_1}{\tau}+2\lambda\mathbf{n}
\end{equation}
and from (\ref{eq:2}) and the definition of $\mathbf{n}$
\begin{equation}
	\lambda=-\frac{|\dot{\mathbf{L}}|^2}{2L}+\frac{\mathbf{w}_2-\mathbf{w}_1}{2\tau}\cdot\mathbf{n}.
\end{equation}
Finally, the equations of motion for the positions of the dumbbell's beads are
\begin{equation}
	\left\{
	\begin{array}{l}
		\dot{\mathbf{v}}_1=-\dfrac{\mathbf{w}_1}{\tau}+\dfrac{|\mathbf{v}_2-\mathbf{v}_1|^2}{2L}\mathbf{n}-\dfrac{\mathbf{w}_2-\mathbf{w}_1}{2\tau}\cdot\mathbf{n}\otimes\mathbf{n} \\[10pt]
		\dot{\mathbf{v}}_2=-\dfrac{\mathbf{w}_2}{\tau}-\dfrac{|\mathbf{v}_2-\mathbf{v}_1|^2}{2L}\mathbf{n}+\dfrac{\mathbf{w}_2-\mathbf{w}_1}{2\tau}\cdot\mathbf{n}\otimes\mathbf{n}.
	\end{array}
	\right.
\end{equation}
The latter equations are essentially the same obtained when imposing no-slip conditions on the two spheres via an immersed boundary method.
If we now take the overdamped (or ${\rm Re}=0$) limit $\tau\to0$, we get
\begin{equation}
	\left\{
	\begin{array}{l}
	0=-\mathbf{w}_1-\dfrac{\mathbf{w}_2-\mathbf{w}_1}{2}\cdot\mathbf{n}\otimes \mathbf{n} \\[10pt]
	0=-\mathbf{w}_2+\dfrac{\mathbf{w}_2-\mathbf{w}_1}{2}\cdot\mathbf{n}\otimes \mathbf{n}. 
	\end{array}
	\right.
	\label{eq:6}
\end{equation}
By summing the two equations one gets $\mathbf{w}_1=-\mathbf{w}_2$, and by substituting this relation into each equation
\begin{equation}
	(\mathbb{I}-\mathbf{n}\otimes \mathbf{n})\mathbf{w}_{1,2}=0.
	\label{eq:7}
\end{equation}
Since $(\mathbf{v}_2-\mathbf{v}_1)\cdot\mathbf{n}=0$ because of inextensibility, we can take the difference of (\ref{eq:7}) for $\mathbf{w}_2$ and $\mathbf{w}_1$ and get
\begin{equation}
	\dot{\mathbf{L}}=(\mathbb{I}-\mathbf{n}\otimes\mathbf{n})(\mathbf{u}_2-\mathbf{u}_1).
\end{equation}
Now, since $\dot{\mathbf{n}}=(\mathbb{I}-\mathbf{n}\otimes \mathbf{n})\dot{\mathbf{L}}/L$, one gets
\begin{equation}
	\dot{\mathbf{n}}=\frac{1}{L}(\mathbb{I}-\mathbf{n}\otimes \mathbf{n})(\mathbb{I}-\mathbf{n}\otimes \mathbf{n})(\mathbf{u}_2-\mathbf{u}_1)=\frac{1}{L}(\mathbb{I}-\mathbf{n}\otimes \mathbf{n})(\mathbf{u}_2-\mathbf{u}_1)
\end{equation}
with the last equality stemming from the idempotence of the projector.
If the dumbbell's length is very small we can write $\mathbf{u}_2-\mathbf{u}_1=\nabla\mathbf{u}{\bf n}L+O(L^2)$, so we get to first order in $L$
\begin{equation}
	\dot{\mathbf{n}}=(\mathbb{I}-{\bf n\otimes n})\nabla\mathbf{u}{\bf n}.
	\label{eq:cvd}
\end{equation}
The latter is Jeffery's equation \cite{jeffery1922motion} with elongation parameter $\Lambda=1$. Indeed Jeffery's equation can be written as
\begin{equation}
	\dot{\mathbf{n}}=\mathbb{O}{\bf n}+\Lambda\mathbb{S}(\mathbb{I}-\mathbf{n}\otimes \mathbf{n}){\bf n}
	\label{jeffery}
\end{equation}
with $\mathbb{O}$ and $\mathbb{S}$ the antisymmetric and symmetric part of the
velocity gradient tensor $\nabla\mathbf{u}$, respectively. Because of symmetry
$\mathbb{O}{\bf n}=(\mathbb{I}-\mathbf{n}\otimes \mathbf{n})\mathbb{O}{\bf n}$ so for $\lambda=1$
one can reconstruct the gradients and obtain (\ref{eq:cvd}). 

\end{document}